\documentclass[conference]{IEEEtran}
\IEEEoverridecommandlockouts
\usepackage{amsmath}
\usepackage{amssymb}
\usepackage{amsfonts}
\usepackage{graphicx}
\usepackage{textcomp}
\usepackage{xcolor}
\usepackage{multirow}
\usepackage{soul}
\usepackage{tikz}
\usepackage{algorithm}
\usepackage{algpseudocode}
\usepackage{colortbl}
\usepackage{bm}
\usepackage{float}
\usepackage{subfigure}
\usepackage{pifont}% http://ctan.org/pkg/pifont

\usepackage{multicol,tabularx,capt-of}
\usepackage{hhline}
\usepackage{multirow}

\newcommand{\cmmnt}[1]{}  

\newcommand\encircle[1]{%
\tikz[baseline=(X.base)] 
  \node (X) [draw, scale=0.75, shape=circle, inner sep=0, fill=black, text=white, minimum size=0em] {\strut #1};}

\newcommand{\cmark}{\ding{51}}%
\newcommand{\xmark}{\ding{55}}%

    \makeatletter 
\newcommand\semiHuge{\@setfontsize\semiHuge{22.72}{27.38}}
\makeatother

\begin{document}

\title{\semiHuge TPU-Gen: LLM-Driven Custom Tensor Processing Unit Generator\vspace{-0.5em}} 
 \author{Deepak Vungarala$^{\dagger}$, Mohammed E. Elbtity$^{\ddagger}$, Sumiya Syed$^{\dagger}$, Sakila Alam$^{\dagger}$, Kartik Pandit$^{\dagger}$,  Arnob Ghosh$^{\dagger}$,\\ Ramtin Zand$^\ddagger$, Shaahin~Angizi$^\dagger$\\
 \small

 $^\dagger$New Jersey Institute of Technology, Newark, NJ, USA,
 $^\ddagger$University of South Carolina, Columbia, SC, USA\\
E-mails: \{dv336,shaahin.angizi\}@njit.edu\\
\vspace{-3.8em}}

\maketitle

\begin{abstract}
The increasing complexity and scale of Deep Neural Networks (DNNs) necessitate specialized tensor accelerators, such as Tensor Processing Units (TPUs), to meet various computational and energy efficiency requirements. Nevertheless, designing optimal TPU remains challenging due to the high domain expertise level, considerable manual design time, and lack of high-quality, domain-specific datasets. This paper introduces TPU-Gen, the first Large Language Model (LLM) based framework designed to automate the exact and approximate TPU generation process, focusing on systolic array architectures. TPU-Gen is supported with a meticulously curated, comprehensive, and open-source dataset that covers a wide range of spatial array designs and approximate multiply-and-accumulate units, enabling design reuse, adaptation, and customization for different DNN workloads. The proposed framework leverages Retrieval-Augmented Generation (RAG) as an effective solution for a data-scare hardware domain in building LLMs, addressing the most intriguing issue, hallucinations. TPU-Gen transforms high-level architectural specifications into optimized low-level implementations through an effective hardware generation pipeline. Our extensive experimental evaluations demonstrate superior performance, power, and area efficiency, with an average reduction in area and power of 92\% and 96\% from the manual optimization reference values. These results set new standards for driving advancements in next-generation design automation tools powered by LLMs. \vspace{-0.5em}
\end{abstract}

% \begin{IEEEkeywords}
% component, formatting, style, styling, insert
% \end{IEEEkeywords}

\section{Introduction}\vspace{-0.4em}

The rising computational demands of Deep Neural Networks (DNNs) have driven the adoption of specialized tensor processing accelerators, such as Tensor Processing Units (TPUs). These accelerators, characterized by low global data transfer, high clock frequencies, and deeply pipelined Processing Elements (PEs), excel in accelerating training and inference tasks by optimizing matrix multiplication \cite{TPU-micro2018}. Despite their effectiveness, the complexity and expertise required for their design remain significant barriers. Static accelerator design tools, such as Gemmini \cite{genc2021gemmini} and DNNWeaver \cite{sharma2016high}, address some of these challenges by providing templates for systolic arrays, data flows, and software ecosystems \cite{ren2023survey,vungarala2023comparative}. However, these tools still face limitations, including complex programming interfaces, high memory usage, and inefficiencies in handling diverse computational patterns \cite{xu2020automatic,angizi2019mrima}. These constraints underscore the need for innovative solutions to streamline hardware design processes. 

Large Language Models (LLMs) have emerged as a promising solution, offering the ability to generate hardware descriptions from high-level design intents. LLMs can potentially reduce the expertise and time required for DNN hardware development by encapsulating vast domain-specific knowledge. However, realizing this potential requires overcoming three critical challenges. First, existing datasets are often limited in size and detail, hindering the generation of reliable designs \cite{chang2023chipgpt,10323953}. Second, while fine-tuning is essential to minimize the human intervention, fine-tuning LLMs often results in hallucinations producing non-sensical or factually incorrect responses, compromising their applicability \cite{thakur2023verigen,jiang2024survey}. Finally, an effective pipeline is needed to mitigate these hallucinations and ensure the generation of consistent, contextually accurate code \cite{jiang2024survey}.
% Also, LLM models suffer from hallucinations resulting in faulty responses which should be mitigated in use cases such as hardware generations as the complexity escalates in hardware \cite{izacard2023atlas}. 
Therefore, the core questions we seek to answer are the following-- \textit{Can there be an effective way to rely on LLM to act as a critical mind and adapt implementations like Retrieval-Augmented Generation (RAG) to minimize hallucinations? Can we leverage domain-specific LLMs with RAG through an effective pipeline to automate the design process of TPU to meet various computational and energy efficiency requirements?}

\begin{table*}[t]
\centering
\caption{Comparison of the Selected LLM-based HDL/HLS generators.} \vspace{-1em}
\scalebox{0.75}{
\begin{tabular}{|l|c|c|c|c|c|c|c|c|c|c|c|c|}
\hline
\rowcolor[HTML]{C0C0C0} 
\multicolumn{1}{|c|}{\cellcolor[HTML]{C0C0C0}\textbf{Property}} & \textbf{Ours} & \textbf{ \cite{thakur2023verigen}} & \textbf{ \cite{10323953}} & \textbf{ \cite{chang2023chipgpt}} & \textbf{ \cite{blocklove2023chip}} & \textbf{ \cite{thakur2023autochip}} & \textbf{\cite{ma2024verilogreader}} & \textbf{\cite{fang2024assertllm}} & \textbf{\cite{liu2024verilogeval}} & \textbf{\cite{zhang2024mg}} &\textbf{\cite{vungarala2024sa}}\\ \hline
Function & TPU Gen.&Verilog Gen.&AI Accel. Gen.&Verilog Gen.&Verilog Gen.&Verilog Gen.& Hardware Verf. & Hardware Verf. &Verilog Gen. &  $\dagger$ & AI Accel. Gen. 
 \\ \hline
Chatbot$^*$                              & \cmark        & \xmark           & \xmark              & \xmark            & \xmark            & \xmark  & \cmark   & \cmark & \xmark    & \xmark  &\xmark   \\ \hline
Dataset                             & \cmark           & \cmark(Verilog)       & \xmark              & NA            & NA            & NA  & \xmark & \xmark    & \cmark & \cmark & \cmark    \\ \hline
Output format                                & Verilog            & Verilog            & HLS             & Verilog & Verilog             & Verilog     & Verilog &HDL & Verilog & Verilog  & Chisel     \\ \hline
Auto. Verif.                      & \cmark              & \xmark            & \xmark              & \xmark            & \xmark             & \cmark   & \cmark & \xmark & \cmark  & \xmark & \cmark     \\ \hline

Human in Loop                           & Low     & Medium           & Medium             & Medium           & High            & Low  & Low & Low & Low & Low   &Low    \\ \hline
Fine tuning   & \cmark & \cmark  & \cmark & \xmark & \xmark & \xmark & \xmark & \xmark & \xmark & \cmark &\xmark  \\ \hline
RAG                     & \cmark          & \xmark           & \xmark             & \xmark          & \xmark            & \xmark  & \xmark   & \cmark   & \xmark & \xmark &\xmark \\
\hline
\end{tabular}}
\label{analysis}
\tiny
$^*$\smash[4]{A user interface featuring Prompt template generation for the input of LLM.} 
$^\dagger$ Not applicable.
\vspace{-3.2em}
\end{table*}

To answer this question, we develop the first-of-its-kind TPU-Gen as an automated exact and approximate
TPU design generation framework with a comprehensive dataset specifically tailored for ever-growing DNN topologies. Our contributions in this paper are threefold: 
(1) Due to the limited availability of annotated data necessary for efficient fine-tuning of an open-source LLM, we introduce a meticulously curated dataset that encompasses various levels of detail and corresponding hardware descriptions, designed to enhance LLMs' learning and generative capabilities in the context of TPU design;
(2) We develop TPU-Gen as a potential solution to reduce hallucinations leveraging RAG and fine-tuning, to align best for the LLMs to streamline the approximate TPU design generation process considering budgetary constraints (e.g., power, latency, area), ensuring a seamless transition from high-level specifications to low-level implementations; and
(3) We design extensive experiments to evaluate our approach's performance and reliability, demonstrating its superiority over existing methods. We anticipate that TPU-Gen will provide a framework that will influence the future trajectory of DNN hardware acceleration research for generations to come\footnote{The dataset and fine-tuned models are open-sourced. The link is omitted to maintain anonymity since the GitHub anonymous link should be under 2GB which is exceeded in this study.}.

\vspace{-0.5em}
\section{background}\vspace{-0.3em}
\textbf{LLM for Hardware Design.} LLMs show promise in generating Hardware Description Language (HDL) and High-Level Synthesis (HLS) code. Table \ref{analysis} compares notable methods in this field. 
% GitHub Copilot \cite{friedman2021introducing} introduces automatic code generation, paving the way for tools like DAVE \cite{pearce2020dave}. 
VeriGen \cite{thakur2023verigen} and ChatEDA \cite{he2023chateda} refine hardware design workflows, automating the RTL to GDSII process with fine-tuned LLMs. ChipGPT \cite{chang2023chipgpt} and Autochip \cite{thakur2023autochip} integrate LLMs to generate and optimize hardware designs, with Autochip producing precise Verilog code through simulation feedback. Chip-Chat \cite{blocklove2023chip} demonstrates interactive LLMs like ChatGPT-4 in accelerating design space exploration. MEV-LLM \cite{nadimi2024multi} proposes multi-expert LLM architecture for Verilog code generation.
RTLLM \cite{lu2023rtllm} and GPT4AIGChip \cite{10323953} enhance design efficiency, showcasing LLMs’ ability to manage complex design tasks and broaden access to AI accelerator design. To the best of our knowledge, GPT4AIGChip \cite{10323953} and SA-DS \cite{vungarala2024sa} are a few initial works focus on an extensive framework specifically aimed at the generation of domain-specific AI accelerator designs where SA-DS focus on creating a dataset in HLS and employ fine-tuning free methods such as single-shot and multi-shot inputs to LLM. Other works for hardware also include creation of SPICE circuits \cite{vungarala2024spicepilot, lai2024analogcoder}. However, the \textit{absence of prompt optimization, tailored datasets, model fine-tuning, and LLM hallucination} pose a barrier to fully harnessing the potential of LLMs in such frameworks \cite{he2023chateda,vungarala2024sa}. This limitation confines their application to standard LLMs without fine-tuning or In-Context Learning (ICL) \cite{he2023chateda}, which are among the most promising methods for optimizing LLMs \cite{dai2022can}. 

\noindent\textbf{Retrieval-Augmented Generation.}
RAG is a promising paradigm that combines deep learning with traditional retrieval techniques to help mitigate hallucinations in LLMs \cite{izacard2023atlas}. RAG leverages external knowledge bases, such as databases, to retrieve relevant information, facilitating the generation of more accurate and reliable responses \cite{chen2309benchmarking, izacard2023atlas}. 
The primary challenge in deploying LLMs for hardware generation or any application lies in their tendency to deviate from the data and hallucinate, making it challenging to capture the essence of circuits and architectural components. LLMs tend to prioritize creativity and finding innovative solutions, which often results in straying from the data \cite{jiang2024survey}. As previous works show, the RAG model can be a cost-efficient solution by retrieving and augmenting data, avoiding heavy computational demands \cite{qin2024robust}.
% It can be implemented using systems like Elasticsearch, where semantic and vector-based searches enhance text sequence generation, enabling models to generate contextually relevant outputs efficiently. Other implementations involve querying external knowledge sources and dynamically modifying them based on user needs, providing computational efficiency and reducing response time. Earlier works combine RAG and LLM to find out the bug in HDL \cite{qayyum2024from}. 

\noindent\textbf{Approximate MAC Units.} Approximate computing has been widely explored as a means to trade reduced accuracy for gains in design metrics, including area, power consumption, and performance \cite{roohi2019apgan,Ansari2021AnIL,angizi2023near,jiang2021non,angizi2018cmp,angizi2018majority}. As the computation core in various PEs in TPUs, several approximate Multiply-and-Accumulate (MAC) units have been proposed as alternatives to precise multipliers and adders and extensively analyzed in accelerating deep learning \cite{Elbtity2020HighSA, Younes2019AlgorithmicLA}. These MAC units are composed of two arithmetic stages—multiplication and accumulation with previous products—each of which can be independently approximated. Most approximate multipliers, such as logarithmic multipliers, are composed of two key components: low-precision arithmetic logic and a pre-processing unit that acts as steering logic to prepare the operands for low-precision computation \cite{Hashemi2015DRUMAD}. These multipliers typically balance accuracy and power efficiency. For example, the logarithmic multiplier introduced in \cite{Ansari2021AnIL} emphasizes accuracy, while the multipliers in \cite{9126271} are designed to reduce power and latency. On the other hand, most approximate adders, such as lower part OR adder (LOA) \cite{dalloo2018systematic}, exploit the fact that extended carry propagation is infrequent, allowing adders to be divided into independent sub-adders shortening the critical path. To preserve computational accuracy, the approximation is applied to the least significant bits of the operands, while the most significant bits remain accurate. \vspace{-1em}

\begin{figure}[t]
\centering
\includegraphics[width=0.72\linewidth]{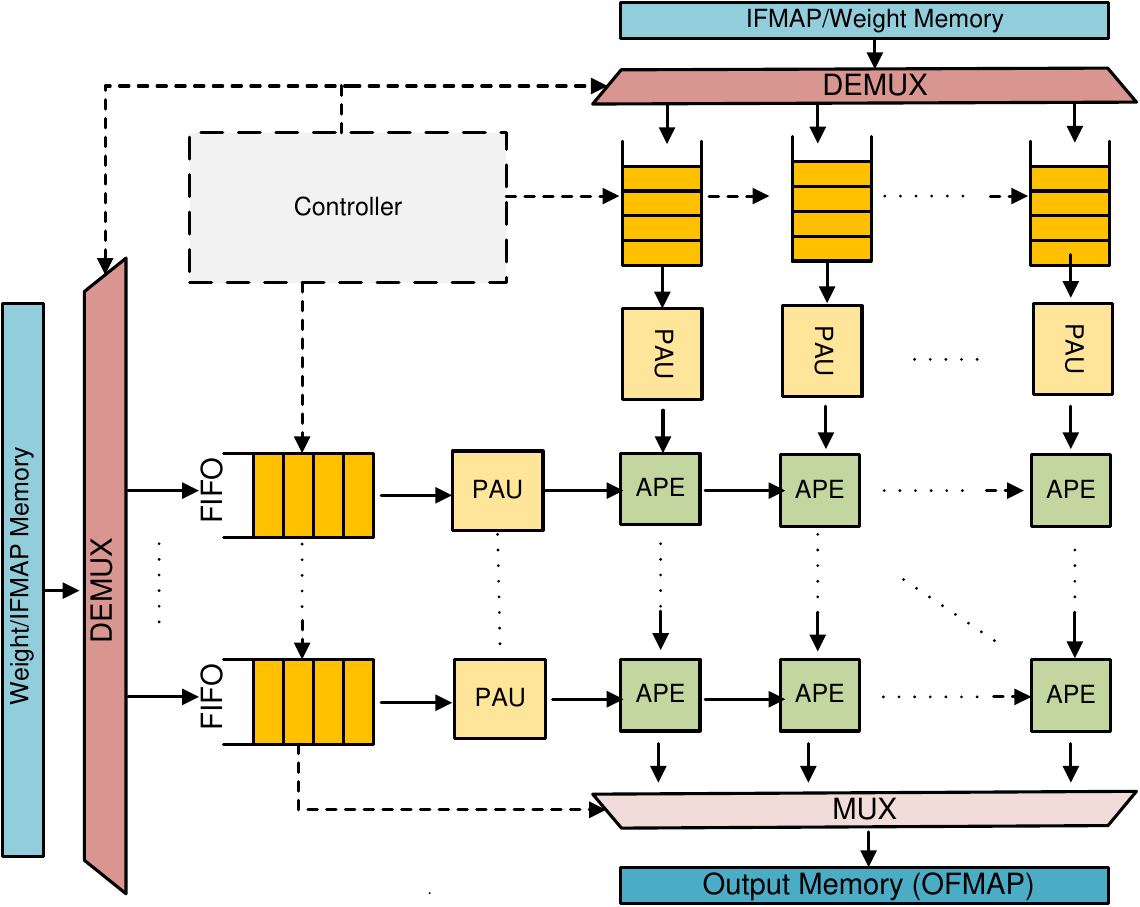}
\vspace{-1em}
\caption{The overall template for TPU design.}
\label{fig:tpu_arch}
\vspace{-1.5em}
\end{figure}

\section{TPU-Gen Framework}
\subsection{Architectural Template}\vspace{-0.5em}
\textbf{Developing a Generic Template.}
The TPU architecture utilizes a systolic array of PEs with MAC units for efficient matrix and vector computations. This design enhances performance and reduces energy consumption by reusing data, minimizing buffer operations \cite{TPU-micro2018}. Input data propagates diagonally through the array in parallel.
The TPU template, illustrated in Fig.~\ref{fig:tpu_arch}, extends the TPU’s systolic array with Output Stationary (OS) dataflow to enable concurrent approximation of input feature maps (IFMaps) and weights. It comprises five components: weight/IFMap memory, FIFOs, a controller, Pre-Approximate Units (PAUs), and Approximate Processing Elements (APEs). The weights and IFMaps are stored in their respective memories, with the controller managing memory access and data transfer to FIFOs per the OS dataflow. PAUs, positioned between FIFOs and APEs, dynamically truncate high-precision operands to lower precision before sending them to APEs, which perform MAC operations using approximate multipliers and adders. Sharing PAUs across rows and columns reduces hardware overhead, introducing minimal latency but significantly improving overall performance \cite{9901385}.

\noindent\textbf{Highly-Parameterized RTL Code.}
We design highly flexible and parameterized RTL codes for 13 different approximate adders and 12 different approximate multipliers as representative approximate circuits. For the approximate adders, we have two tunable parameters: the bit-width and the imprecise part. The bit-width specifies the number of bits for each operand and the imprecise part specifies the number of inexact bits in the adder output. For the approximate multipliers, we have one common parameter, i.e., Width (W), which specifies the bit-width of the multiplication operands. We also have more tunable parameters based on specific multipliers, some of which are listed in Table \ref{tab:mult_param}.
\begin{table}
    \centering
    %\small
    \caption{Approximate multiplier hyper-parameters} \vspace{-1em}
    \scalebox{0.7}{\begin{tabular}{|l|c|c|c|}
    \hline
        \cellcolor[HTML]{C0C0C0}\textbf{Design} & \cellcolor[HTML]{C0C0C0}\textbf{Parameter} & \cellcolor[HTML]{C0C0C0}\textbf{Description} & \cellcolor[HTML]{C0C0C0}\textbf{Default} \\
        \hline
        \multirow{1}{*}{BAM \cite{Farshchi2013NewAM}} & VBL & No. of zero bits during partial product generation & W/2 \\
        \hline
        \multirow{1}{*}{ALM\_LOA \cite{Liu2018DesignAE}} & M & Inaccurate part of LOA adder & W/2  \\
        \hline
        \multirow{1}{*}{ALM\_MAA3 \cite{Liu2018DesignAE}} & M & Inaccurate part of MAA3 adder & W/2 \\
        \hline
        \multirow{1}{*}{ALM\_SOA \cite{Liu2018DesignAE}} & M & Inaccurate part of SOA adder & W/2 \\
        \hline
        \multirow{1}{*}{ASM \cite{Sarwar2018EnergyEfficientNC}} & Nibble\_Width & number of precomputed alphabets & 4\\
        \hline
        \multirow{1}{*}{DRALM \cite{9126271}} & MULT\_DW & Truncated bits of each operand & W/2 \\
        \hline
        \multirow{1}{*}{RoBA \cite{Zendegani2017RoBAMA}} & ROUND\_WIDTH & Scales the widths of the shifter & 1 \\
        \hline
    \end{tabular}} \vspace{-2.5em}
    \label{tab:mult_param}
\end{table}
%\end{sidewaystable}
%\end{table*}
We leveraged the parametrized RTL library of approximate arithmetic circuits to build a TPU library that enables automatic selection of the systolic array size $S$, bit precision $n$, and one of the approximate multipliers and approximate adders. The internal parameters that are used to tune the approximate arithmetic libraries are also included in the TPU parameterized RTL library, thus, allowing the user to have complete flexibility to adjust their designs to meet specific hardware specifications and application accuracy requirements. Moreover, we developed a design automation methodology, enabling the automatic implementation and simulation of many TPU circuits in various simulation platforms such as Design Compiler and Vivado. In addition to the highly parameterized RTL codes, we developed TCL and Python scripts to autonomously measure their error, area, performance, and power dissipation under various constraints. \vspace{-0.5em}

\subsection{Framework Overview}  
\begin{figure}[t] %\vspace{-1em}
    \centering
    \includegraphics[width=1\linewidth]{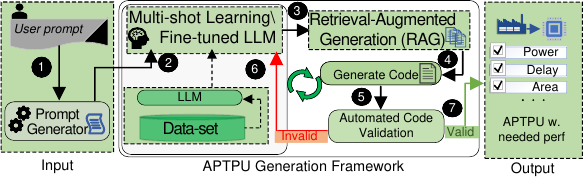} \vspace{-2.5em}
    \caption{The proposed TPU-Gen framework.} \vspace{-2em}
    \label{framefig}
\end{figure}

\begin{figure}[b] \vspace{-1em} 
    \centering 
\includegraphics[width=1.01\linewidth]{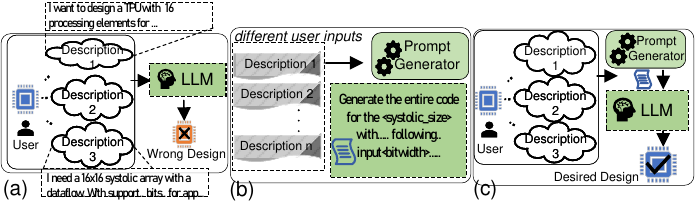}  \vspace{-2em} % This can be changed to a strip rather this grid approach, we can discuss on this.
    \caption{(a) Multiple descriptions for a single TPU design demonstrate that a design can be verbally defined in numerous ways, potentially misleading LLMs in generating the intended design, (b)  Proposed prompt generator extracts the required features from the given verbal descriptions, (c) Using a script to generate a verbal description aligned with the training data.}
    \label{4fig}\vspace{-1.5em}
\end{figure}

\vspace{-0.2em}
TPU-Gen framework depicted in Fig. \ref{framefig} targets the development of domain-specific LLMs, emphasizing the interplay between the model's responses and two key factors: the input prompt and the model's learned parameters. The framework optimizes both elements to enhance LLM's performance. An initial prompt conveying the user's intent and key software and hardware specifications of the intended TPU design and application is enabled through the \textit{Prompt Generator} in Step \encircle{1}. A verbal description of a tensor processing accelerator design can often result in a many-to-one mapping as shown in Fig. \ref{4fig}(a), especially when such descriptions do not align with the format of the training dataset. This misalignment increases the likelihood of hallucinations in the LLM's output, potentially leading to faulty designs \cite{niu2024mitigating}. To minimize hallucinations and incorrect outputs in LLM-generated designs, studies have shown that inputs adhering closely to patterns observed in the training data produce more accurate and desirable results \cite{zhang2024mg,vungarala2024sa}. However, this critical aspect has often been overlooked in previous state-of-the-art research \cite{10323953}, with some researchers opting instead to address the issue through prompt optimization techniques \cite{vungarala2024sa}. In this framework, we tackle the problem by employing a script that extracts key features, such as systolic size and relevant metrics, from any given verbal input by the user. These features are then embedded into a template, which serves as the prompt for the LLM input. As a domain-specific LLM, TPU-Gen focuses on generating the most valuable RTL top file detailing the circuit, and blocks involved in the presented architectural template in Section III.A.

An immediate usage of the proposed dataset explained in Section III.C in TPU-Gen is to help fine-tune a generic LLM for the task of TPU design, where the input with a prompt will be fed to the LLM (Step \encircle{2} in Fig. \ref{framefig}). Equivalently, one may employ ICL, or multi-shot learning as a more computationally efficient compromise to fine-tuning \cite{dai2022can}. The multi-shot prompting techniques can be used where the proposed dataset will function as the source for multi-shot examples. Given that the TPU-Gen dataset integrates verbal descriptions with corresponding TPU systolic array design pairs, the LLM generates a TPU's top-level file as the output in Verilog. This top-level file includes all necessary architectural module dependencies to ensure a fully functional design (step \encircle{3}). Further, we propose to leverage the RAG module to generate the other dependency files into the project, completing the design (step \encircle{4}).
Next, a third-party quality evaluation tool can be employed to provide a quantitative evaluation of the design, verify functional correctness, and integrate the design with the full stack (step \encircle{5}). Here, for quality and functional evaluation, the generated designs, initially described in Verilog, are synthesized using YOSYS \cite{Yosys}. This synthesis process incorporates an automated RTL-to-GDSII validation stage, where the generated designs are evaluated and classified as either \textit{Valid} or \textit{Invalid} based on the completeness of their code sequences and the correctness of their input-output relationships. \textit{Valid} designs proceed to resource validation, where they are optimized with respect to Power, Performance, and Area (PPA) metrics. In contrast, designs flagged as \textit{Invalid} initiate a feedback loop for error analysis and subsequent LLM retraining, enabling iterative refinement (steps \encircle{2} to \encircle{6}) to achieve predefined performance criteria. Ultimately, designs that successfully pass these stages in step \encircle{7} are ready for submission to the foundry. \vspace{-0.7em}
 
\begin{figure}[b] %\vspace{-1em}
    \centering
    \includegraphics[width=0.95\linewidth]{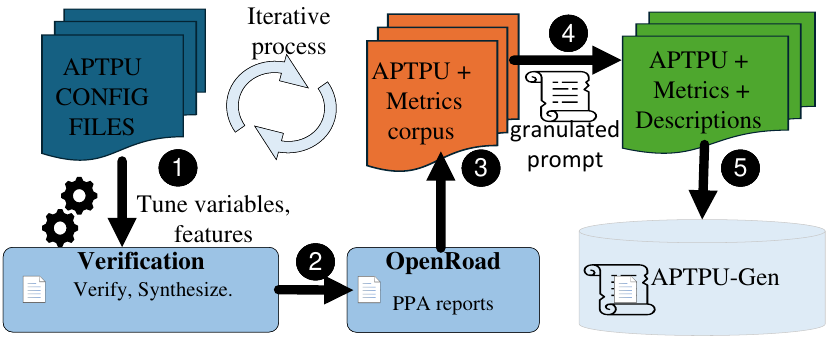} \vspace{-1em}
    \caption{TPU-Gen dataset curation.} \vspace{-2em}
    \label{aptpu}
\end{figure}

\subsection{Dataset Curation} \vspace{-0.4em}
Leveraging the parameterized RTL code of the TPU, we develop a script to systematically explore various architectural configurations and generate a wide range of designs within the proposed framework (step \encircle{1} in Fig. \ref{aptpu}). The generated designs undergo synthesis and functional verification (step \encircle{2}). Subsequently, the OpenROAD suite \cite{OPENROAD} is employed to produce PPA metrics (step \encircle{3}). The PPA data is parsed using Pyverilog (step \encircle{4}), resulting in the creation of a detailed, multi-level dataset that captures the reported PPA metrics (step \encircle{5}).
Steps \encircle{1} to \encircle{3} are iterated until all architectural variations are generated. The time required for each data point generation varies depending on the specific configuration. To efficiently populate the TPU-Gen dataset, we utilize multiple scripts that automate the generation of data points across different systolic array sizes, ensuring comprehensive coverage of design space exploration. Fig. \ref{aptpu} shows the detailed methodology underpinning our dataset creation. The validation when compared to prior works \cite{thakur2023verigen,pearce2020dave} understanding we work in a different design space abstraction makes it tough to have a fair comparison. However, looking by the scale of operation and the framework's efficiency we require minimal efforts comparatively.

\begin{figure}[t] \vspace{-2em}
    \centering
\includegraphics[width=0.9\linewidth,height=4.2cm]{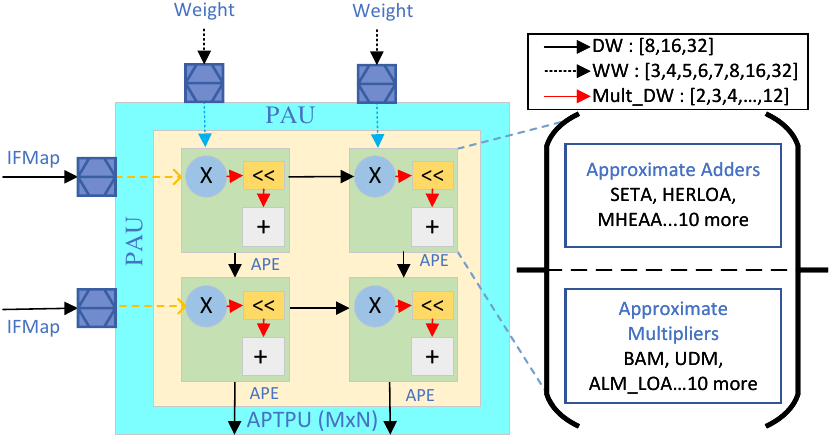}\vspace{-1.2em}
    \caption{An example of one category and its design space parameters.} \vspace{-1em}
    \label{aptpu1}
\end{figure}

Fig. \ref{aptpu1} visualizes the selection of different circuits to make PAUs and APEs accommodating different input Data Widths (DW) (8, 16, 32 bits) and Weight Widths (WW) (ranging from 3 to 32 bits) to generate approximate MAC units. 
These feature units highlight the flexible template of the TPU and enhance its adaptability and performance across various DNN workloads. Including lower bit-width weights is particularly advantageous for highly quantified models, enabling efficient processing with reduced computational resources.

\begin{figure}[b] \vspace{-1em}
    \centering
\includegraphics[width=1\linewidth]{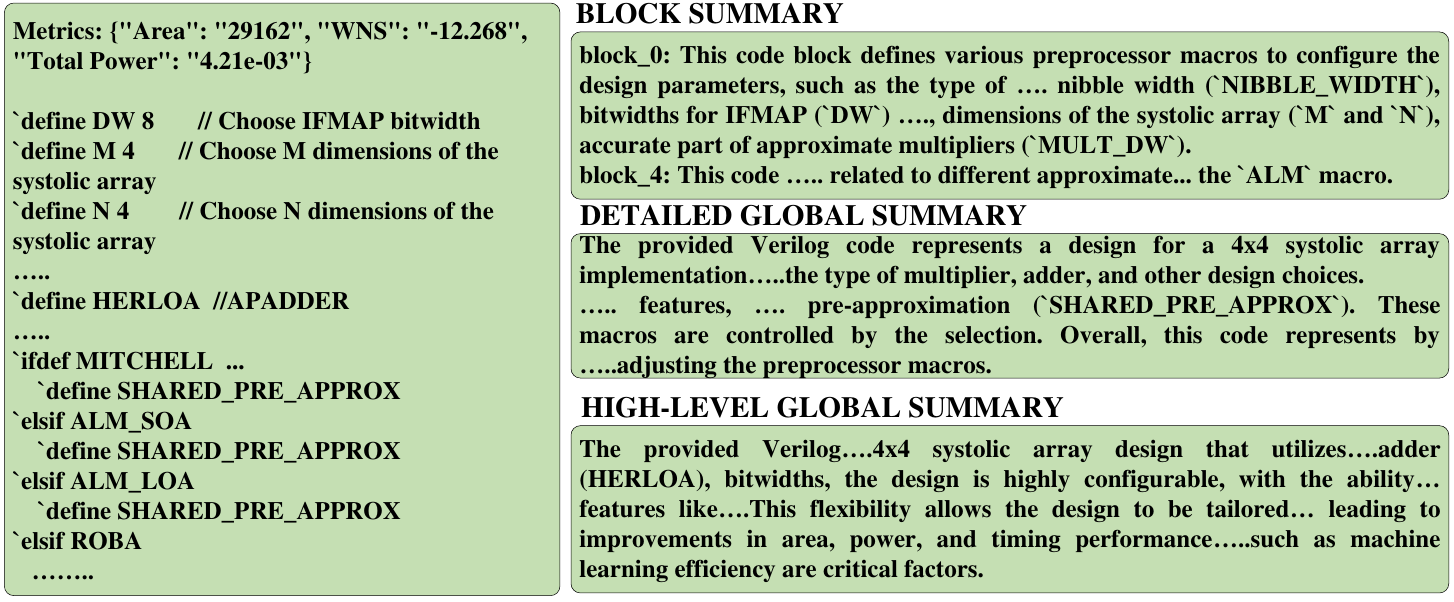}
\vspace{-2em}
    \caption{An example of a data point by adapting MG-V format.}
    \label{dpstruct}\vspace{-1em}
\end{figure}

TPU-Gen dataset offers 29,952 possible variations for a systolic array size with 8 different systolic array implementations to facilitate various workloads spanning from 4$\times$4 for smaller loads to 256$\times$256 to crunch bigger DNN workloads. Accounting for the systolic size variations in the TPU-Gen dataset promises a total of 29,952$\times$8 = 2,39,616 data points with PPA metrics reported. While TPU-Gen is constantly growing with newer data points, we checkpoint our dataset creation currently reported as having 25,000 individual TPU designs. We provide two variations: $(i)$ A top module file consisting of details of the entire circuit implementation, which can be used in cases such as RAG implementation to save the computation resources, and $(ii)$ A detailed, multi-level granulated dataset, as depicted in Fig. \ref{dpstruct}, is curated by adapting MG-Verilog \cite{zhang2024mg} to assist LLM in generating Verilog code to support the development of a highly sophisticated, fine-tuned model. This model facilitates the automated generation of individual hardware modules, intelligent integration, deployment, and reuse across various designs and architectures. Please note that due to the domain-specific nature of the dataset, some data redundancy is inevitable, as similar modules are reused and reconfigured to construct new TPUs with varying architectural configurations. This structured dataset enables efficient exploration and customization of TPU designs while ensuring that the generated modules can be systematically adapted for different design requirements, leading to enhanced flexibility and scalability in hardware design automation. Additionally, we provide detailed metrics for each design iteration, which aid the LLM in generating budget-constrained designs or in creating an efficient design space exploration strategy to accelerate the result optimization process. \vspace{-0.32em}
%Every data point for APTPU-Gen, as exemplified in Fig. \ref{dpstruct}, reflects the modifications described in previous sections where we incorporated MG-Verilog \cite{zhang2024mg} as supplementary metadata. 

\begin{table*}[t] \vspace{-2em}
\centering
\caption{Prompts to successfully generate exact TPU modules via TPU-Gen.} \vspace{-1em}
\label{tb2}
\scalebox{0.96}{
\begin{tabular}{|ll|cccc|cccc|}
\hline
\rowcolor[HTML]{C0C0C0} 
\multicolumn{2}{|c|}{\cellcolor[HTML]{C0C0C0}} & \multicolumn{4}{c|}{\cellcolor[HTML]{C0C0C0}\textbf{Module Generation}} & \multicolumn{4}{c|}{\cellcolor[HTML]{C0C0C0}\textbf{Module Integration}} \\ \cline{3-10} 
\rowcolor[HTML]{C0C0C0} 
\multicolumn{2}{|c|}{\multirow{-2}{*}{\cellcolor[HTML]{C0C0C0}\textbf{LLM Model}}} & \multicolumn{1}{c|}{\cellcolor[HTML]{C0C0C0}\textbf{Pass@1}} & \multicolumn{1}{c|}{\cellcolor[HTML]{C0C0C0}\textbf{Pass@3}} & \multicolumn{1}{c|}{\cellcolor[HTML]{C0C0C0}\textbf{Pass@5}} & \textbf{Pass@10} & \multicolumn{1}{c|}{\cellcolor[HTML]{C0C0C0}\textbf{Pass@1}} & \multicolumn{1}{c|}{\cellcolor[HTML]{C0C0C0}\textbf{Pass@3}} & \multicolumn{1}{c|}{\cellcolor[HTML]{C0C0C0}\textbf{Pass@5}} & \textbf{Pass@10} \\ \hline
\multicolumn{2}{|l|}{\textbf{Mistral-7B (Q3)}} & \multicolumn{1}{c|}{17\%} & \multicolumn{1}{c|}{83\%} & \multicolumn{1}{c|}{100\%} & 100\% & \multicolumn{1}{c|}{0\%} & \multicolumn{1}{c|}{25\%} & \multicolumn{1}{c|}{75\%} & 100\% \\ \hline
\multicolumn{2}{|l|}{\textbf{CodeLlama-7B (Q4)}} & \multicolumn{1}{c|}{0\%} & \multicolumn{1}{c|}{50\%} & \multicolumn{1}{c|}{83\%} & 100\% & \multicolumn{1}{c|}{0\%} & \multicolumn{1}{c|}{50\%} & \multicolumn{1}{c|}{75\%} & 100\% \\ \hline
\multicolumn{2}{|l|}{\textbf{CodeLlama-13B (Q4)}} & \multicolumn{1}{c|}{66\%} & \multicolumn{1}{c|}{83\%} & \multicolumn{1}{c|}{100\%} & 100\% & \multicolumn{1}{c|}{25\%} & \multicolumn{1}{c|}{75\%} & \multicolumn{1}{c|}{100\%} & 100\% \\ \hline
\multicolumn{2}{|l|}{\textbf{Claude 3.5 Sonnet}} & \multicolumn{1}{c|}{83\%} & \multicolumn{1}{c|}{100\%} & \multicolumn{1}{c|}{100\%} & 100\% & \multicolumn{1}{c|}{75\%} & \multicolumn{1}{c|}{100\%} & \multicolumn{1}{c|}{100\%} & 100\% \\ \hline
\multicolumn{2}{|l|}{\textbf{ChatGPT-4o}} & \multicolumn{1}{c|}{83\%} & \multicolumn{1}{c|}{100\%} & \multicolumn{1}{c|}{100\%} & 100\% & \multicolumn{1}{c|}{50\%} & \multicolumn{1}{c|}{100\%} & \multicolumn{1}{c|}{100\%} & 100\% \\ \hline
\multicolumn{2}{|l|}{\textbf{Gemmini Advanced}} & \multicolumn{1}{c|}{50\%} & \multicolumn{1}{c|}{50\%} & \multicolumn{1}{c|}{74\%} & 91\% & \multicolumn{1}{c|}{25\%} & \multicolumn{1}{c|}{75\%} & \multicolumn{1}{c|}{74\%} & 91\% \\ \hline
\end{tabular}
} \vspace{-1em}
\end{table*}

\section{Experiment Results} \vspace{-0.5em}
\subsection{Objectives}\vspace{-0.5em}
\par We designed four distinct experiments employing various approaches, each tailored to the unique capabilities of LLMs such as GPT \cite{openai2024gpt4o}, Gemini \cite{Gemini}, and Claude \cite{Claude}, as well as the best open-source models from the leader board \cite{leaderboard}. Each model is deployed in experiments aligning with the study's objectives and anticipated outcomes. Experiments 1 focus on observing the prompting mechanism that assists LLM in generating the desired output by implementing ICL; with this knowledge, we develop the prompt template discussed in Sections III-B. Experiment 2 focuses on adapting the proposed TPU-Gen framework by fine-tuning LLM models. For fine-tuning, we used 4$\times$A100 GPU with 80GB VRAM. Experiment 3 is to demonstrate the effectiveness of RAG in TPU-Gen and it's applicability for hardware design. 
Experiment 4 tests the TPU-Gen framework's ability to generate designs efficiently with an industry-standard 45nm technology library. Throughout the process, we also consider hardware under the given PPA budget to ensure the feasibility of achieving the objectives outlined in the initial phases. 

\subsection{Experiments and Results}\vspace{-0.4em}

\begin{figure} [b] \vspace{-1em}
    \centering
    \includegraphics[width=1.01\linewidth]{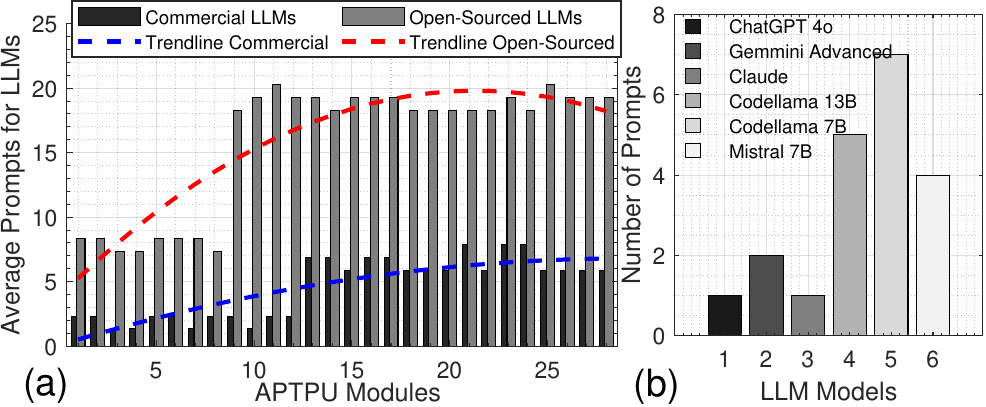}\vspace{-1em}
    \caption{Average TPU-Gen prompts for (a) Module Generation, and (b) Module Integration via LLMs.} \vspace{-1.5em}
    \label{exp-plot_aptpu}
\end{figure}

\subsubsection{Experiment 1: ICL-Driven TPU Generation and Approximate Design Adaptation.}
We evaluate the capability of LLMs to generate and synthesize a novel TPU architecture and its approximate version using TPU-Gen. Utilizing the prompt template from \cite{vungarala2024sa}, we refined it to harness LLM capabilities better. LLM performance is assessed on two metrics: $(i)$~\textbf{Module Generation}—the ability to generate required modules, and $(ii)$~\textbf{Module Integration}—the capability to construct the top module by integrating components. We tested commercial models like \cite{openai2024gpt4o, Gemini} via chat interfaces and open-source models listed in Table~\ref{tb2}, using LM Studio \cite{lmstudio}. For the TPU, we successfully developed the design and obtained the GDSII layout (Fig.~\ref{tp-layout}(a)). Commercial models performed well with a single prompt at pass@1, averaging 72\% in module generation and 50\% in integration. Open-source models performed better with the increase of pass@k, averaging 72\% for pass@1 in module generation to 100\% and 50\% to 100\% upscale from pass@3 to pass@10 in integration. For the approximate TPU, involving approximate circuit algorithms, we provided example circuits and used ICL and Chain of Thought (CoT) to guide the LLMs. Open-source models struggled due to a lack of specialized knowledge, as shown in Fig.~\ref{exp-plot_aptpu}. The design layout from this experiment is in Fig.~\ref{tp-layout}(b). All outputs were manually verified using test benches. This is the first work to generate both exact and approximate TPU architectures using prompting to LLM. However, significant human expertise and intervention are required, especially for complex architectures like approximate circuits. To minimize the human involvement, we implement fine-tuning.

\vspace{0.5em}
\hspace{-0.5em}\fcolorbox{black}{white}{\begin{minipage}{24em}
{\small \textbf{Takeaway 1.} \textit{LLMs with efficient prompting are capable of generating exact and approximate TPU modules and integrate them to create complete designs. However, human involvement is extensively required, especially for novel architectures. Fine-tuning LLMs is necessary to reduce human intervention and facilitate the exploration of new designs.}}
\end{minipage}}

\subsubsection{Experiment 2:  Full TPU-Gen Implementation}\vspace{1em}
This experiment investigates cost-efficient approaches for adapting domain-specific language models to hardware design. In previous experiments, we observed that limited spatial and hierarchical hardware knowledge hindered LLM performance in integrating circuits. The TPU-Gen template (Fig. \ref{framefig}) addresses this by delegating creative tasks to the LLM and retrieving dependent modules via RAG, optimizing AI accelerator design while reducing computational overhead and minimizing LLM hallucinations.
ICL experiments show that fine-tuning enhances LLM reliability. The TPU-Gen proposes a way to develop domain-specific LLMs with minimal data. The experiment used a TPU-Gen dataset version 1 of 5,000 Verilog headers DW and WW inputs. This dataset comprises systolic array implementations with biased approximate circuit variations. We split data statically in 80:20 for training and testing open-source LLMs \cite{leaderboard}, with two primary goals of $1.$ Analyzing the impact of the prompt template generator on the fine-tuned LLM's performance  (Table \ref{lasttable}). 
 $2.$ Investigating the RAG model for hardware development.\vspace{-0.35em}

 \begin{figure}[b]\vspace{-1em}
    \centering
    \includegraphics[width=0.95\linewidth]{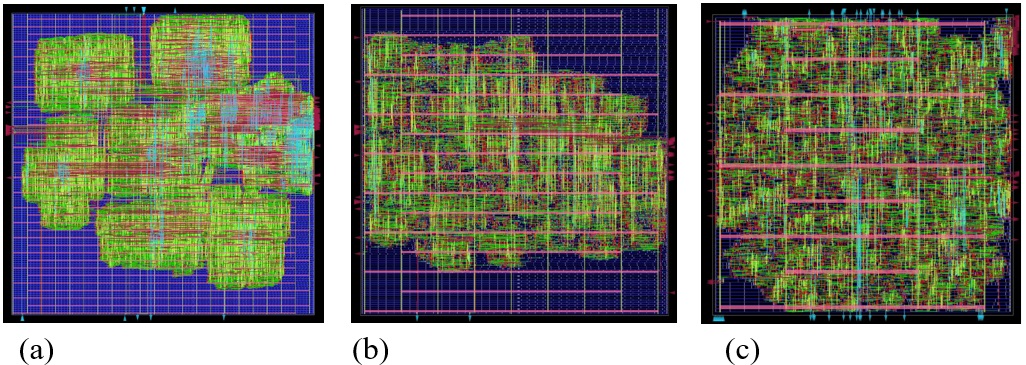} \vspace{-1em}
    \caption{A GDSII layout of (a) TPU, (b) TPU by prompting LLM, (c) approximate TPU by TPU-Gen framework.}
    \label{tp-layout}\vspace{-1.5em}
\end{figure}

All models used Low-Rank Adaptation (LoRA) fine-tuning with the Adam optimizer at a learning rate of $1e^{-5}$. The fine-tuned models were evaluated to generate the desired results efficiently with a random prompt at pass@$1$ to generate the TPU. From Table \ref{lasttable}, we can observe that the outputs without the prompt generator are labeled as failures as they were unsuitable for further development and RAG integration. We can observe the same prompt when parsed to the prompt-template generator with a single try; we score an accuracy of 86.6\%.  Further, we used RAG and then processed the generated Verilog headers for module retrieval. According to \cite{jiang2024survey}, LLMs tend to prioritize creativity and finding innovative solutions, which often results in straying from the data. To address this, we employed a compute and cost-efficient method. This shows that the fine-tuning along with RAG can greatly enhance the performance. 
Fig. \ref{tp-layout}(c) shows the GDSII layout of the design generated by the TPU-Gen framework. 

\begin{table}[t]
\centering \vspace{-1.5em}
\caption{Prompt Generator vs Human inputs to Fine-tuned models.} \vspace{-1em}
\scalebox{0.92}{
\label{lasttable}
\begin{tabular}{|c|cc|cc|}
\hline
\rowcolor[HTML]{C0C0C0} 
\cellcolor[HTML]{C0C0C0} & \multicolumn{2}{c|}{\cellcolor[HTML]{C0C0C0}\textbf{Prompt Template}} & \multicolumn{2}{c|}{\cellcolor[HTML]{C0C0C0}\textbf{Human Input}} \\ \cline{2-5} 
\rowcolor[HTML]{C0C0C0} 
\multirow{-2}{*}{\cellcolor[HTML]{C0C0C0}\textbf{Models}} & \multicolumn{1}{c|}{\cellcolor[HTML]{C0C0C0}\textbf{Pass}} & \textbf{Fail} & \multicolumn{1}{c|}{\cellcolor[HTML]{C0C0C0}\textbf{Pass}} & \textbf{Fail} \\ \hline
\multicolumn{1}{|l|}{\textbf{CodeLlama-7B-hf}} & \multicolumn{1}{c|}{27} & 03 & \multicolumn{1}{c|}{01} & 29 \\ \hline
\textbf{CodeQwen1.5 -7B}& \multicolumn{1}{c|}{25} & 05 & \multicolumn{1}{c|}{0} & 30 \\ \hline
\textbf{Mistral -7B} & \multicolumn{1}{c|}{28} & 02 & \multicolumn{1}{c|}{02} & 28 \\ \hline
\textbf{Starcoder2-7B} & \multicolumn{1}{c|}{24} & 06 & \multicolumn{1}{c|}{0} & 30 \\ \hline
\end{tabular}} \vspace{-2.5em}
\end{table}

\vspace{0.5em}
\hspace{-0.5em}\fcolorbox{black}{white}{\begin{minipage}{24em}
{\small \textbf{Takeaway 2.} \textit{Prompting techniques such as prompt template steer LLM to generate desired results after fine-tuning, as observed 86\% success in generation. RAG, a cost-efficient method to generate the hardware modules reliably, completing the entire Verilog design for an application with minimal computational overhead.}
}
\end{minipage}} \vspace{0.2em}
\subsubsection{Experiment 3: Significance of RAG}
To assess the effectiveness of RAG in the TPU-Gen framework, we evaluated 1,000 Verilog header codes generated by fine-tuned LLMs under two conditions: with and without RAG integration. Table~\ref{tab:ragnorag} presents results over 30 designs tested by our framework to generate complete project files. Without RAG, failures occurred due to output token limitations and hallucinated variables. RAG is essential as the design is not a standalone file to compile. Validated header codes were provided in the RAG-enabled pipeline, and required modules were dynamically retrieved from the RAG database, ensuring fully functional and accurate designs. Conversely, models without RAG relied solely on internal knowledge, leading to hallucinations, token constraints, and incomplete designs. Models using RAG consistently achieved pass rates exceeding 95\%, with Mistral-7B and CodeLlama-7B-hf attaining 100\% success. In contrast, all models failed entirely without RAG, underscoring its pivotal role in ensuring design accuracy and addressing LLM limitations. RAG provides a robust solution to key challenges in fine-tuned LLMs for TPU hardware design by retrieving external information from the RAG database, ensuring contextual accuracy, and significantly reducing hallucinations. Additionally, RAG dynamically fetches dependencies in a modular manner, enabling the generation of complete and accurate designs without exceeding token limits. RAG is a promising solution in this context since our models were fine-tuned with only Verilog header data detailing design features. However, fine-tuning models with the entire design data would expose LLMs to severe hallucinations and token limitations, making generating detailed and functional designs challenging.\vspace{-1.5em}

\begin{table}[h] 
%\vspace{-2em}
\centering
\caption{significance of RAG in TPU-Gen.} 
\vspace{-1em}
\label{tab:ragnorag}
\scalebox{0.96}{
\begin{tabular}{|ll|cc|cc|}
\hline
\rowcolor[HTML]{C0C0C0} 
\multicolumn{2}{|c|}{\cellcolor[HTML]{C0C0C0}} & \multicolumn{2}{c|}{\cellcolor[HTML]{C0C0C0}\textbf{With RAG}} & \multicolumn{2}{c|}{\cellcolor[HTML]{C0C0C0}\textbf{Without RAG}} \\ \cline{3-6} 
\rowcolor[HTML]{C0C0C0} 
\multicolumn{2}{|c|}{\multirow{-2}{*}{\cellcolor[HTML]{C0C0C0}\textbf{LLM Model}}} & \multicolumn{1}{c|}{\cellcolor[HTML]{C0C0C0}\textbf{Pass(\%)}} & \textbf{Fail(\%)} & \multicolumn{1}{c|}{\cellcolor[HTML]{C0C0C0}\textbf{Pass(\%)}} & \textbf{Fail(\%)} \\ \hline
\multicolumn{2}{|l|}{\textbf{CodeLama-7B-hf}} & \multicolumn{1}{c|}{100} & 0 & \multicolumn{1}{c|}{0} & 100 \\ \hline
\multicolumn{2}{|l|}{\textbf{Mistral-7B}} & \multicolumn{1}{c|}{100} & 0 & \multicolumn{1}{c|}{0} & 100 \\ \hline
\multicolumn{2}{|l|}{\textbf{CodeQwen1.5-7B}} & \multicolumn{1}{c|}{95} & 5 & \multicolumn{1}{c|}{0} & 100 \\ \hline
\multicolumn{2}{|l|}{\textbf{StarCoder2-7B}} & \multicolumn{1}{c|}{98} & 2 & \multicolumn{1}{c|}{0} & 100 \\ \hline
\end{tabular}
} 
\end{table}

\hspace{-0.5em}\fcolorbox{black}{white}{\begin{minipage}{24em}
{\small \textbf{Takeaway 3.} \textit{The experiment highlights the significance of the RAG usage with a fine-tuned model to avoid hallucinations and let LLM be creative consistently.}
}
\end{minipage}} \vspace{0.5em}

\subsubsection{Experiment 4: Design Generation Efficiency} 
\begin{figure}[t] \vspace{-0.5em}
    \centering
\includegraphics[width=1\linewidth]{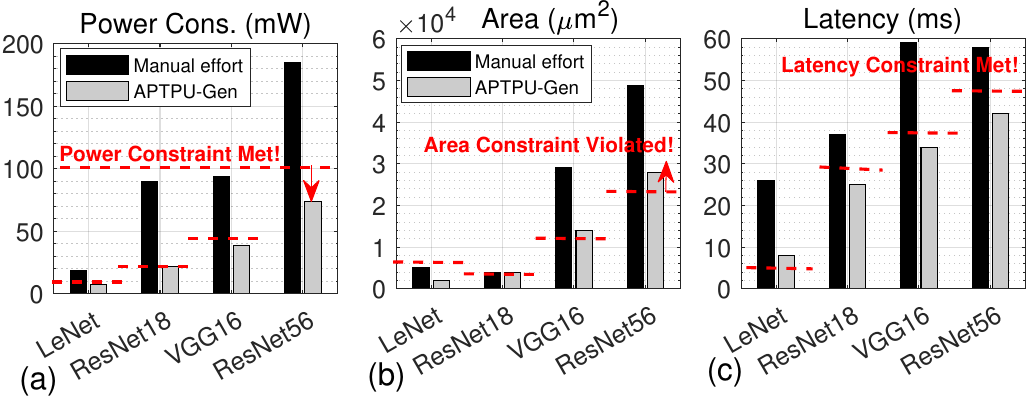} \vspace{-2em}
    \caption{PPA metrics comparison for TPU architectures generated by TPU-Gen and the manual user: (a) Power consumption, (b) Area, (c) Latency.}
    \label{analysis2}\vspace{-1.25em}
\end{figure}

Building on the successful generation of approximate TPU in experiment 2, here we evaluate and benchmark the architectures produced by the TPU-Gen framework as the work performed in this paper is the first of it's kind we are comparing against manual optimization created by expert human designers, focusing on power, area, and latency as shown in Fig. \ref{analysis2}(a)-(c). We utilize four DNN architectures for this evaluation: LeNet, ResNet18, VGG16, and ResNet56, performing inference tasks on the MNIST, CIFAR-10, SVHN, and CIFAR-100 datasets. In the manually optimized designs, a skilled hardware engineer fine-tunes parameters within the TPU template. This iterative optimization process is repeated until no further performance gains can be achieved within a reasonable timeframe of approximately one day \cite{10323953}, or the expert determines, based on empirical results, that additional refinements would yield minimal benefits. Using the PPA metrics as reference values (e.g., 100mW, 0.25mm$^2$, 48ms for ResNet56), both TPU-Gen and the manual user are tasked with generating the TPU architecture. Fig. \ref{analysis2} illustrates that across a range of network architectures, TPU-Gen consistently yields results with minimal deviation from the reference benchmarks. In contrast, the manual designs exhibit significant violations in terms of PPA.

\vspace{0.75em}
\hspace{-0.5em}\fcolorbox{black}{white}{\begin{minipage}{24em}
{\small \textbf{Takeaway 4.} \textit{TPU-Gen consistently yields results with minimal deviation from the PPA reference, whereas the manual designs exhibit significant violations.}}
\end{minipage}}

\section{Conclusions}
This paper introduces TPU-Gen, a novel dataset and a novel framework for TPU generation, addressing the complexities of generating AI accelerators amidst rapid AI model evolution. A key challenge, hallucinated variables, is mitigated using an RAG approach, dynamically adapting hardware modules. RAG enables cost-effective, full-scale RTL code generation, achieving budget-constrained outputs via fine-tuned models. Our extensive experimental evaluations demonstrate superior performance, power, and area efficiency, with an average reduction in area and power of 92\% and 96\% from the manual optimization reference values. These results set new standards for driving advancements in next-generation design automation tools powered by LLMs. We are committed to releasing the dataset and fine-tuned models publicly if accepted.

\bibliographystyle{IEEEtran}
\bibliography{Reference}\vspace{-2em}

% Generated by IEEEtran.bst, version: 1.14 (2015/08/26)
\begin{thebibliography}{10}
\providecommand{\url}[1]{#1}
\csname url@samestyle\endcsname
\providecommand{\newblock}{\relax}
\providecommand{\bibinfo}[2]{#2}
\providecommand{\BIBentrySTDinterwordspacing}{\spaceskip=0pt\relax}
\providecommand{\BIBentryALTinterwordstretchfactor}{4}
\providecommand{\BIBentryALTinterwordspacing}{\spaceskip=\fontdimen2\font plus
\BIBentryALTinterwordstretchfactor\fontdimen3\font minus \fontdimen4\font\relax}
\providecommand{\BIBforeignlanguage}[2]{{%
\expandafter\ifx\csname l@#1\endcsname\relax
\typeout{** WARNING: IEEEtran.bst: No hyphenation pattern has been}%
\typeout{** loaded for the language `#1'. Using the pattern for}%
\typeout{** the default language instead.}%
\else
\language=\csname l@#1\endcsname
\fi
#2}}
\providecommand{\BIBdecl}{\relax}
\BIBdecl

\bibitem{TPU-micro2018}
N.~Jouppi, C.~Young, N.~Patil, and D.~Patterson, ``Motivation for and evaluation of the first tensor processing unit,'' \emph{IEEE Micro}, vol.~38, no.~3, pp. 10--19, 2018.

\bibitem{genc2021gemmini}
H.~Genc \emph{et~al.}, ``Gemmini: Enabling systematic deep-learning architecture evaluation via full-stack integration,'' in \emph{2021 58th ACM/IEEE Design Automation Conference (DAC)}.\hskip 1em plus 0.5em minus 0.4em\relax IEEE, 2021, pp. 769--774.

\bibitem{sharma2016high}
H.~Sharma, J.~Park, D.~Mahajan, E.~Amaro, J.~K. Kim, C.~Shao, A.~Mishra, and H.~Esmaeilzadeh, ``From high-level deep neural models to fpgas,'' in \emph{2016 49th Annual IEEE/ACM International Symposium on Microarchitecture (MICRO)}.\hskip 1em plus 0.5em minus 0.4em\relax IEEE, 2016, pp. 1--12.

\bibitem{ren2023survey}
W.-Q. Ren \emph{et~al.}, ``A survey on collaborative dnn inference for edge intelligence,'' \emph{Machine Intelligence Research}, vol.~20, no.~3, pp. 370--395, 2023.

\bibitem{vungarala2023comparative}
D.~Vungarala, M.~Morsali, S.~Tabrizchi, A.~Roohi, and S.~Angizi, ``Comparative study of low bit-width dnn accelerators: Opportunities and challenges,'' in \emph{2023 IEEE 66th International Midwest Symposium on Circuits and Systems (MWSCAS)}.\hskip 1em plus 0.5em minus 0.4em\relax IEEE, 2023, pp. 797--800.

\bibitem{xu2020automatic}
P.~Xu and Y.~Liang, ``Automatic code generation for rocket chip rocc accelerators,'' 2020.

\bibitem{angizi2019mrima}
S.~Angizi, Z.~He, A.~Awad, and D.~Fan, ``Mrima: An mram-based in-memory accelerator,'' \emph{IEEE Transactions on Computer-Aided Design of Integrated Circuits and Systems}, vol.~39, no.~5, pp. 1123--1136, 2019.

\bibitem{chang2023chipgpt}
K.~Chang, Y.~Wang, H.~Ren, M.~Wang, S.~Liang, Y.~Han, H.~Li, and X.~Li, ``Chipgpt: How far are we from natural language hardware design,'' \emph{arXiv preprint arXiv:2305.14019}, 2023.

\bibitem{10323953}
Y.~Fu, Y.~Zhang, Z.~Yu, S.~Li, Z.~Ye, C.~Li, C.~Wan, and Y.~C. Lin, ``Gpt4aigchip: Towards next-generation ai accelerator design automation via large language models,'' in \emph{2023 IEEE/ACM International Conference on Computer Aided Design (ICCAD)}.\hskip 1em plus 0.5em minus 0.4em\relax IEEE, 2023, pp. 1--9.

\bibitem{thakur2023verigen}
S.~Thakur, B.~Ahmad, H.~Pearce, B.~Tan, B.~Dolan-Gavitt, R.~Karri, and S.~Garg, ``Verigen: A large language model for verilog code generation,'' \emph{ACM Transactions on Design Automation of Electronic Systems}, vol.~29, no.~3, pp. 1--31, 2024.

\bibitem{jiang2024survey}
X.~Jiang, Y.~Tian, F.~Hua, C.~Xu, Y.~Wang, and J.~Guo, ``A survey on large language model hallucination via a creativity perspective,'' \emph{arXiv preprint arXiv:2402.06647}, 2024.

\bibitem{blocklove2023chip}
J.~Blocklove, S.~Garg, R.~Karri, and H.~Pearce, ``Chip-chat: Challenges and opportunities in conversational hardware design,'' in \emph{2023 ACM/IEEE 5th Workshop on Machine Learning for CAD (MLCAD)}.\hskip 1em plus 0.5em minus 0.4em\relax IEEE, 2023, pp. 1--6.

\bibitem{thakur2023autochip}
S.~Thakur, J.~Blocklove, H.~Pearce, B.~Tan, S.~Garg, and R.~Karri, ``Autochip: Automating hdl generation using llm feedback,'' \emph{arXiv preprint arXiv:2311.04887}, 2023.

\bibitem{ma2024verilogreader}
R.~Ma, Y.~Yang, Z.~Liu, J.~Zhang, M.~Li, J.~Huang, and G.~Luo, ``Verilogreader: Llm-aided hardware test generation,'' \emph{arXiv:2406.04373v1}, 2024.

\bibitem{fang2024assertllm}
W.~Fang \emph{et~al.}, ``Assertllm: Generating and evaluating hardware verification assertions from design specifications via multi-llms,'' \emph{arXiv:2402.00386v1}, 2024.

\bibitem{liu2024verilogeval}
M.~Liu, N.~Pinckney, B.~Khailany, and H.~Ren, ``Verilogeval: Evaluating large language models for verilog code generation,'' \emph{arXiv:2309.07544v2}, 2024.

\bibitem{zhang2024mg}
Y.~Zhang, Z.~Yu, Y.~Fu, C.~Wan, and Y.~C. Lin, ``Mg-verilog: Multi-grained dataset towards enhanced llm-assisted verilog generation,'' \emph{arXiv preprint arXiv:2407.01910}, 2024.

\bibitem{vungarala2024sa}
D.~Vungarala, M.~Nazzal, M.~Morsali, C.~Zhang, A.~Ghosh, A.~Khreishah, and S.~Angizi, ``Sa-ds: A dataset for large language model-driven ai accelerator design generation,'' \emph{arXiv e-prints}, pp. arXiv--2404, 2024.

\bibitem{he2023chateda}
H.~Wu \emph{et~al.}, ``Chateda: A large language model powered autonomous agent for eda,'' \emph{IEEE Transactions on Computer-Aided Design of Integrated Circuits and Systems}, 2024.

\bibitem{nadimi2024multi}
B.~Nadimi and H.~Zheng, ``A multi-expert large language model architecture for verilog code generation,'' \emph{arXiv preprint arXiv:2404.08029}, 2024.

\bibitem{lu2023rtllm}
Y.~Lu, S.~Liu, Q.~Zhang, and Z.~Xie, ``Rtllm: An open-source benchmark for design rtl generation with large language model,'' in \emph{2024 29th Asia and South Pacific Design Automation Conference (ASP-DAC)}.\hskip 1em plus 0.5em minus 0.4em\relax IEEE, 2024, pp. 722--727.

\bibitem{vungarala2024spicepilot}
D.~Vungarala, S.~Alam, A.~Ghosh, and S.~Angizi, ``Spicepilot: Navigating spice code generation and simulation with ai guidance,'' \emph{arXiv preprint arXiv:2410.20553}, 2024.

\bibitem{lai2024analogcoder}
Y.~Lai, S.~Lee, G.~Chen, S.~Poddar, M.~Hu, D.~Z. Pan, and P.~Luo, ``Analogcoder: Analog circuit design via training-free code generation,'' \emph{arXiv preprint arXiv:2405.14918}, 2024.

\bibitem{dai2022can}
D.~Dai, Y.~Sun, L.~Dong, Y.~Hao, S.~Ma, Z.~Sui, and F.~Wei, ``Why can gpt learn in-context? language models implicitly perform gradient descent as meta-optimizers,'' \emph{arXiv preprint arXiv:2212.10559}, 2022.

\bibitem{izacard2023atlas}
G.~Izacard \emph{et~al.}, ``Atlas: Few-shot learning with retrieval augmented language models,'' \emph{Journal of Machine Learning Research}, vol.~24, no. 251, pp. 1--43, 2023.

\bibitem{chen2309benchmarking}
J.~Chen, H.~Lin, X.~Han, and L.~Sun, ``Benchmarking large language models in retrieval-augmented generation,'' \emph{arXiv preprint arXiv:2309.01431}, 2023.

\bibitem{qin2024robust}
R.~Qin \emph{et~al.}, ``Robust implementation of retrieval-augmented generation on edge-based computing-in-memory architectures,'' \emph{arXiv:2405.04700v1}, 2024.

\bibitem{roohi2019apgan}
A.~Roohi, S.~Sheikhfaal, S.~Angizi, D.~Fan, and R.~F. DeMara, ``Apgan: Approximate gan for robust low energy learning from imprecise components,'' \emph{IEEE Transactions on Computers}, vol.~69, no.~3, pp. 349--360, 2019.

\bibitem{Ansari2021AnIL}
M.~S. Ansari, B.~Cockburn, and J.~Han, ``An improved logarithmic multiplier for energy-efficient neural computing,'' \emph{IEEE Trans. on Comput.}, vol.~70, pp. 614--625, 2021.

\bibitem{angizi2023near}
S.~Angizi, M.~Morsali, S.~Tabrizchi, and A.~Roohi, ``A near-sensor processing accelerator for approximate local binary pattern networks,'' \emph{IEEE Transactions on Emerging Topics in Computing}, vol.~12, no.~1, pp. 73--83, 2023.

\bibitem{jiang2021non}
H.~Jiang, S.~Angizi, D.~Fan, J.~Han, and L.~Liu, ``Non-volatile approximate arithmetic circuits using scalable hybrid spin-cmos majority gates,'' \emph{IEEE Transactions on Circuits and Systems I: Regular Papers}, vol.~68, no.~3, pp. 1217--1230, 2021.

\bibitem{angizi2018cmp}
S.~Angizi, Z.~He, A.~S. Rakin, and D.~Fan, ``Cmp-pim: an energy-efficient comparator-based processing-in-memory neural network accelerator,'' in \emph{Proceedings of the 55th Annual Design Automation Conference}, 2018, pp. 1--6.

\bibitem{angizi2018majority}
S.~Angizi, H.~Jiang, R.~F. DeMara, J.~Han, and D.~Fan, ``Majority-based spin-cmos primitives for approximate computing,'' \emph{IEEE Transactions on Nanotechnology}, vol.~17, no.~4, pp. 795--806, 2018.

\bibitem{Elbtity2020HighSA}
\BIBentryALTinterwordspacing
M.~E. Elbtity, H.-W. Son, D.-Y. Lee, and H.~Kim, ``High speed, approximate arithmetic based convolutional neural network accelerator,'' \emph{2020 International SoC Design Conference (ISOCC)}, pp. 71--72, 2020. [Online]. Available: \url{https://api.semanticscholar.org/CorpusID:231826033}
\BIBentrySTDinterwordspacing

\bibitem{Younes2019AlgorithmicLA}
H.~Younes, A.~Ibrahim, M.~Rizk, and M.~Valle, ``Algorithmic level approximate computing for machine learning classifiers,'' \emph{2019 26th IEEE Int. Conf. on Electron., Circuits and Syst. (ICECS)}, pp. 113--114, 2019.

\bibitem{Hashemi2015DRUMAD}
S.~Hashemi, R.~I. Bahar, and S.~Reda, ``{DRUM}: A dynamic range unbiased multiplier for approximate applications,'' \emph{2015 IEEE/ACM Int. Conf. on Comput.-Aided Design (ICCAD)}, pp. 418--425, 2015.

\bibitem{9126271}
P.~Yin, C.~Wang, H.~Waris, W.~Liu, Y.~Han, and F.~Lombardi, ``Design and analysis of energy-efficient dynamic range approximate logarithmic multipliers for machine learning,'' \emph{IEEE Transactions on Sustainable Computing}, vol.~6, no.~4, pp. 612--625, 2021.

\bibitem{dalloo2018systematic}
A.~Dalloo, A.~Najafi, and A.~Garcia-Ortiz, ``Systematic design of an approximate adder: The optimized lower part constant-or adder,'' \emph{IEEE Transactions on Very Large Scale Integration (VLSI) Systems}, vol.~26, no.~8, pp. 1595--1599, 2018.

\bibitem{9901385}
M.~E. Elbtity, P.~S. Chandarana, B.~Reidy, J.~K. Eshraghian, and R.~Zand, ``Aptpu: Approximate computing based tensor processing unit,'' \emph{IEEE Transactions on Circuits and Systems I: Regular Papers}, vol.~69, no.~12, pp. 5135--5146, 2022.

\bibitem{Farshchi2013NewAM}
F.~Farshchi \emph{et~al.}, ``New approximate multiplier for low power digital signal processing,'' \emph{The 17th CSI International Symposium on Computer Architecture \& Digital Systems (CADS 2013)}, pp. 25--30, 2013.

\bibitem{Liu2018DesignAE}
W.~Liu \emph{et~al.}, ``Design and evaluation of approximate logarithmic multipliers for low power error-tolerant applications,'' \emph{IEEE Trans. on Circuits and Syst. I: Reg. Papers}, vol.~65, pp. 2856--2868, 2018.

\bibitem{Sarwar2018EnergyEfficientNC}
S.~S. Sarwar \emph{et~al.}, ``Energy-efficient neural computing with approximate multipliers,'' \emph{ACM Journal on Emerging Technologies in Computing Systems (JETC)}, vol.~14, pp. 1 -- 23, 2018.

\bibitem{Zendegani2017RoBAMA}
\BIBentryALTinterwordspacing
R.~Zendegani \emph{et~al.}, ``Roba multiplier: A rounding-based approximate multiplier for high-speed yet energy-efficient digital signal processing,'' \emph{IEEE Transactions on Very Large Scale Integration (VLSI) Systems}, vol.~25, pp. 393--401, 2017. [Online]. Available: \url{https://api.semanticscholar.org/CorpusID:206810935}
\BIBentrySTDinterwordspacing

\bibitem{niu2024mitigating}
M.~Niu, H.~Li, J.~Shi, H.~Haddadi, and F.~Mo, ``Mitigating hallucinations in large language models via self-refinement-enhanced knowledge retrieval,'' \emph{arXiv preprint arXiv:2405.06545}, 2024.

\bibitem{Yosys}
\BIBentryALTinterwordspacing
(2024) Yosys. [Online]. Available: \url{https://github.com/YosysHQ/yosys}
\BIBentrySTDinterwordspacing

\bibitem{OPENROAD}
\BIBentryALTinterwordspacing
(2018) Openroad. [Online]. Available: \url{https://github.com/The-OpenROAD-Project/OpenROAD}
\BIBentrySTDinterwordspacing

\bibitem{pearce2020dave}
H.~Pearce \emph{et~al.}, ``Dave: Deriving automatically verilog from english,'' in \emph{MLCAD}, 2020, pp. 27--32.

\bibitem{openai2024gpt4o}
\BIBentryALTinterwordspacing
(2024) Openai gpt-4. [Online]. Available: \url{https://openai.com/index/hello-gpt-4o/}
\BIBentrySTDinterwordspacing

\bibitem{Gemini}
\BIBentryALTinterwordspacing
(2024) Gemini. [Online]. Available: \url{https://deepmind.google}
\BIBentrySTDinterwordspacing

\bibitem{Claude}
\BIBentryALTinterwordspacing
(2023) Anthropic. [Online]. Available: \url{https://www.anthropic.com}
\BIBentrySTDinterwordspacing

\bibitem{leaderboard}
Evalplus leaderboard. \url{https://evalplus.github.io/leaderboard.html}. Accessed: 2024-09-21.

\bibitem{lmstudio}
``Lm studio - discover, download, and run local llms,'' \url{https://lmstudio.ai/}, accessed: 2024-09-21.

\end{thebibliography}
\end{document}